%% file: nips2016.tex
\pdfoutput=1
\documentclass{article}

\usepackage[final, nonatbib]{nips_2016}

\usepackage[utf8]{inputenc} 
\usepackage[T1]{fontenc}    
\usepackage{url}            
\usepackage{booktabs}       
\usepackage{amsfonts}       
\usepackage{nicefrac}       
\usepackage{microtype}      

\usepackage{times}
\usepackage{graphicx}
\usepackage{subfig}
\usepackage{color}

\usepackage{amssymb}
\usepackage{amsmath}

\title{Towards End-to-End Audio-Sheet-Music Retrieval}

\author{
  Matthias Dorfer, Andreas Arzt and Gerhard Widmer \\
  Department of Computational Perception \\
  Johannes Kepler University Linz \\
  Altenberger Str. 69, A-4040 Linz \\
  \texttt{matthias.dorfer@jku.at} \\
}

\begin{document}

\maketitle

\begin{abstract}
%
%
This paper demonstrates the feasibility of learning to retrieve short snippets
of sheet music (images) when given a short query excerpt of music (audio)
-- and vice versa --, without any symbolic representation of music or
scores. This would be highly useful in many content-based musical retrieval
scenarios. Our approach is based on Deep Canonical Correlation
Analysis (DCCA) and learns correlated latent spaces allowing for
cross-modality retrieval in both directions. Initial experiments
with relatively simple monophonic music show promising results.
\end{abstract}

%
\section{Introduction}
\label{sec:introduction}
\input{introduction}

%
\section{Methods}
\label{sec:methods}
\input{methods}

%
\section{Experiments}
\label{sec:experiments}
\input{experiments}

%
\section{Conclusion}
\label{sec:conclusion}
\input{conclusion}

\subsubsection*{Acknowledgments}
This work is supported by the Austrian Ministries BMVIT and BMWFW,
and the Province of Upper Austria via the COMET Center SCCH,
and by the European Research Council (ERC Grant Agreement
670035, project CON ESPRESSIONE).
The Tesla K40 used for this research was donated by the NVIDIA
corporation.

\small
\bibliographystyle{plain}
\bibliography{nips2016_ws}

\end{document}

%% file: introduction.tex
\begin{figure}[h]
 \centerline{\includegraphics[width=0.8\columnwidth]{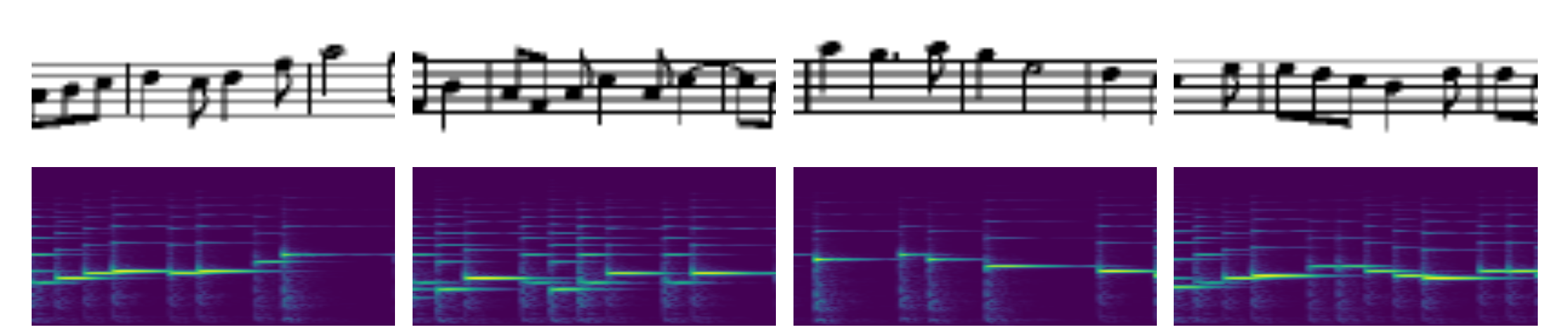}}
 \caption{Example of the data considered for audio to sheet image retrieval.
 		   Top row: short snippets of sheet music images.
 		   Bottom row: Spectrogram excerpts of the corresponding audio (music).}
 \label{fig:correspondence_samples}
\end{figure}

Efficient systems for content-based music retrieval allow for browsing, exploring and managing large music collections. In this paper we tackle the problem of \emph{audio to sheet music matching}, i.e. matching short snippets of music (audio) and corresponding parts in the sheet music (image). Amongst other applications, this is especially useful for large-scale music digitisation projects \cite{Fremerey2008automatic}, which collect large numbers of sheet music and performances and need to link this data to each other.
%
%
It was recently shown that convolutional neural networks
are suitable for dealing with images of sheet music
applied to the task of score following \cite{Dorfer2016Towards}.
Inspired by \cite{Dorfer2016Towards} and \cite{Yan2015DeepCorr},
who showed that DCCA can  efficiently be used to match image and text data,
we propose an end-to-end neural network approach that allows for the retrieval
of short snippets of sheet music (images)
when given a short query excerpt of music (audio).
Figure \ref{fig:correspondence_samples} shows some examples of audio-sheet correspondences
targeted in the present work.

%

%% file: methods.tex
We first introduce a common notation used throughout the paper
and review the concepts of classic and Deep Canonical Correlation Analysis (DCCA) \cite{Andrew2013DCCA}.
Based on DCCA we show how we use it in our system to retrieve
the corresponding sheet image snippet for a given query audio fragment
and vice versa.

Let ${\mathbf{x}_1, ..., \mathbf{x}_N} = \mathbf{X} \in \mathbb{R}^{N \times d_x}$
and ${\mathbf{y}_1, ..., \mathbf{y}_N} = \mathbf{Y} \in \mathbb{R}^{N \times d_y}$
denote a set of $N$ multi-view observations.
Here $\mathbf{X}$ refers to the set of sheet music (score) snippets
and $\mathbf{Y}$ to the corresponding set of audio (spectrogram) snippets
(compare Figure \ref{fig:correspondence_samples}).
Following \cite{Wang_2015_MultiView}, we define $\mathbf{f}$ and $\mathbf{g}$
to be non-linear feature mappings used for processing the raw input data.
In our application we implement $\mathbf{f}$ and $\mathbf{g}$ as two different
convolutional neural networks producing hidden feature representations
$\mathbf{f}(\mathbf{X}) \in \mathbb{R}^{N \times h}$ and
$\mathbf{g}(\mathbf{Y}) \in \mathbb{R}^{N \times h}$ for their corresponding input views.
The parameters of the two models are referred to as
$\Theta_\mathbf{f}$ and $\Theta_\mathbf{g}$.
As in \cite{Yan2015DeepCorr, Andrew2013DCCA} the dimensionality $h$
of the topmost hidden representations is defined to be the same for both views.
We also denote $\mathbf{f}(\mathbf{X})$ and $\mathbf{g}(\mathbf{Y})$
by $\mathbf{f}_X$ and $\mathbf{g}_Y$ respectively for a briefer notation in the reminder of the paper.

Our audio to sheet image retrieval approach is based on (D)CCA, a method from classic multivariate statistics
that relies on the covariance structures of the respective input (latent) feature distributions.
Equation (\ref{eq:Sigmas}) introduces the covariance matrices
for the learned feature representations of both views.
\begin{equation}
\label{eq:Sigmas}
\mathbf{\Sigma}_X = \frac{1}{N-1} \bar{\mathbf{f}}_X^T \bar{\mathbf{f}}_X \; \text{and} \;
\mathbf{\Sigma}_Y = \frac{1}{N-1} \bar{\mathbf{g}}_Y^T \bar{\mathbf{g}}_Y
\end{equation}
In addition to the individual covariance matrices,
CCA requires the cross-covariance $\mathbf{\Sigma}_{XY}$
between the features of the two different views:
\begin{equation}
\label{eq:SigmaXY}
\mathbf{\Sigma}_{XY} = \frac{1}{N-1} \bar{\mathbf{f}}_X^T \bar{\mathbf{g}}_Y
\end{equation}

\subsection{Deep Canonical Correlation Analysis (DCCA)}
In \cite{Andrew2013DCCA}, a deep neural network extension to classical CCA
is introduced for combining the topmost feature representations of
two different neural networks $\mathbf{f}$ and $\mathbf{g}$.
The DCCA optimization target pushes the networks to learn highly correlated feature representations.
Based on the covariances introduced above CCA defines a matrix $\mathbf{T} = \mathbf{\Sigma}_X^{-1/2} \mathbf{\Sigma}_{XY} \mathbf{\Sigma}_Y^{-1/2}$.
The total correlation between $\mathbf{f}_X$ and $\mathbf{g}_Y$ is then computed
as the sum over the singular values $\mathbf{d}$
with corresponding singular value problem $\mathbf{T}=\mathbf{U}\mathbf{D}\mathbf{V}$
and $\mathbf{D} = diag(\mathbf{d})$.
$\mathbf{U}$ and $\mathbf{V}$ are the projection matrices
which transform the two views into the linear CCA sub-space.
The correlation itself is optimized by maximizing the sum over
the singular values $\mathbf{d}$ with respect to the
network parameters $\Theta_\mathbf{f}$ and $\Theta_\mathbf{g}$:
\begin{equation}
\label{eq:cca_obj}
\underset{\Theta_\mathbf{f}, \Theta_\mathbf{g}}{\arg \max} \sum_{i=1}^{h}d_i
\end{equation}
If $\mathbf{f}$ and $\mathbf{g}$ have the same feature dimensionality $h$
it is also possible to optimize the canonical correlation
by maximizing the matrix trace norm $||\mathbf{T}||_{tr}=tr((\mathbf{T}^T \mathbf{T})^{1/2})$.
For a detailed derivation of the DCCA optimization target we refer to \cite{Andrew2013DCCA}.

\subsection{Deep Canonical Correlated Audio-Sheet-Music Retrieval}
\begin{figure}[t!]
 \centerline{\includegraphics[width=0.8\columnwidth]{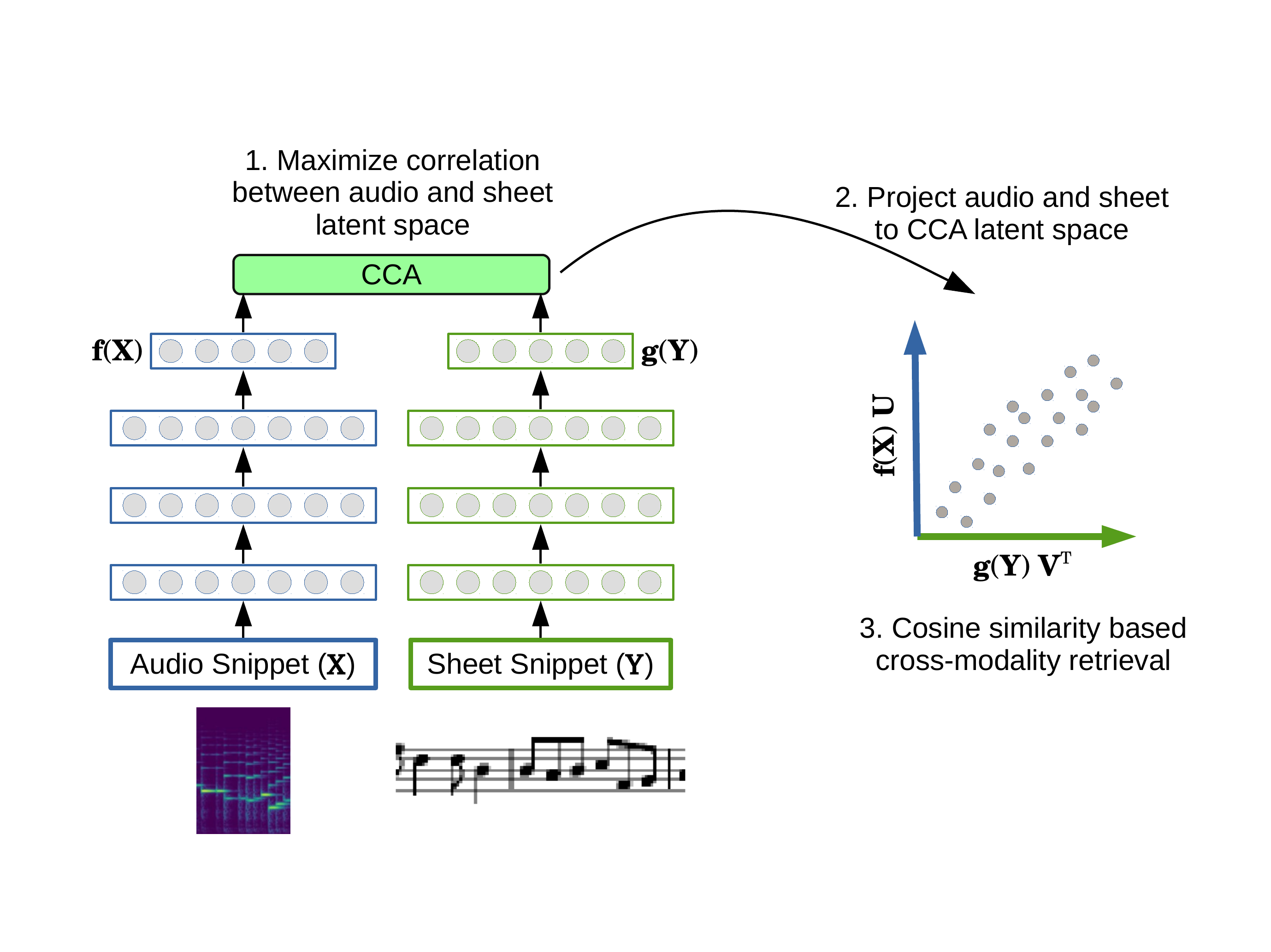}}
 \vspace{-28px}
 \caption{Overview of the audio-to-sheet cross-modality retrieval framework.}
 \label{fig:concept_sketch}
\end{figure}

The proposed audio-sheet-music cross-modality retrieval model
is built on top of two paths of convolutional neural networks.
Both networks operate directly on the respective input modality
and reduce its dimensionality to an $h$-dimensional latent representation.
Figure \ref{fig:concept_sketch} shows a schematic sketch of the entire retrieval pipeline.
Once the network and the corresponding CCA model are trained,
the input data is projected by $\mathbf{f}'_X=\mathbf{f}_X \mathbf{U}$
and $\mathbf{g}'_Y = \mathbf{g}_Y \mathbf{V}^T$ into the CCA space
(with normalized projection matrices $\mathbf{U} \leftarrow \mathbf{\Sigma}_X^{-1/2} \mathbf{U}$
and $\mathbf{V} \leftarrow \mathbf{\Sigma}_Y^{-1/2} \mathbf{V}$).
$\mathbf{f}'_X$ is the projection of a sheet image snippet $\mathbf{x}$ and
$\mathbf{g}'_Y$ is the projection of the corresponding audio snippet $\mathbf{y}$.
A beneficial property of the CCA projection space is that
if a set of pairs exhibits high correlation
then the individual pairs also have a low cosine distance \cite{Yan2015DeepCorr}.
One can exploit this property for retrieval by cosine distance computation
\begin{equation}
d_{cos} = 1.0 - \frac{\mathbf{f}'_X \cdot \mathbf{g}'_Y}{|\mathbf{f}'_X||\mathbf{g}'_Y|}
\end{equation}
e.g. of a query audio vector $\mathbf{g}'_Y$ to a database of reference image vectors $\{\mathbf{f}'_X\}_M$
where $M$ is the number of available candidate sheet image snippets.
The result is a ranking of sheet image snippets
and allows for a selection of the snippet with highest
cross-modality similarity (e.g. lowest cosine distance).
The database of image snippets is thereby created
and processed by the image network $\mathbf{f}$ prior to retrieval time.
This further means that we know for each (indexed) sheet image snippet
(1) the originating piece as well as the
(2) respective sheet image position.
The procedure described above works analogously in the opposite direction
for retrieving audio from given query sheet images.

%% file: experiments.tex
We run our experiments on the same dataset
that was used by \cite{Dorfer2016Towards}
for evaluating their end-to-end score following system in sheet music images.
We further describe our network architectures
as well as the optimization strategies
and introduce the quantitative measures used for evaluation.

\subsection{Data and Experimental Setup}
As in \cite{Dorfer2016Towards} we consider the Nottingham piano midi dataset for our experiments.
The dataset is a collection of midi files split into train, validation and test set.
In terms of data preparation we follow \cite{Dorfer2016Towards} and
(1) render the midi files to sheet music images using Lilypond \footnote{http://www.lilypond.org},
(2) synthesize the midi files to audio and
(3) establish correspondences between short snippets of sheet music and their corresponding excerpt of audio.
For audio preparation the only pre-processing step is computing
log-spectrograms with a sample rate of 22.05kHz,
a FFT window size of 2048, and a computation rate of 31.25 frames per second.
These spectrograms (136 frequency bins) are then directly fed into
the audio part of our cross-modality network.
Figure \ref{fig:correspondence_samples} shows a set of audio-to-sheet correspondences
presented to our network for training.
One audio excerpt comprises $100$ frames and the dimension
of the sheet image snippet is $40 \times 100$ pixel.

The parameters of our model are optimized using stochastic gradient descent with momentum
using a batch size of $100$,
an initial learning rate of $0.1$ and a fixed momentum of $0.9$.
The learning rate is halved every $25$ epochs during training.
Table \ref{tab:model_architecture} provides details on our retrieval architecture.
Our model is basically a VGG style \cite{simonyan2014very} network consisting of sequences of $3 \times 3$ convolution stacks followed by $2 \times 2$ max pooling.
As activations we use Exponential Linear Units (ELUs) \cite{Clevert2015ELU} for all
layers except for the final layer before DCCA where no non-linearity is used at all.
For reducing the dimensionality to the desired correlation space dimensionality $h$ (in our case 32)
we insert as a final building block a $1 \times 1$ convolution having $h$ feature maps
followed by global-average-pooling \cite{LinCY2013NIN}.
The output ($\mathbf{f}_X$ and $\mathbf{g}_Y$) of these layers is than fed into the DCCA optimization target.
\begin{table*}[ht]
\begin{center}
\begin{tabular}{c|c}
\hline
Sheet-Image $40 \times 100$ & Spectrogram $136 \times 100$ \\
\hline
$2\times$Conv($3$, pad-1)-$16$-BN-ELU + MP($2$)		&$2\times$ Conv($3$, pad-1)-$16$-BN-ELU + MP($2$) \\
$2\times$Conv($3$, pad-1)-$32$-BN-ELU + MP($2$)		&$2\times$ Conv($3$, pad-1)-$32$-BN-ELU + MP($2$) \\
$2\times$Conv($3$, pad-1)-$64$-BN-ELU + MP($2$)		&$2\times$ Conv($3$, pad-1)-$64$-BN-ELU + MP($2$) \\
$2\times$Conv($3$, pad-1)-$64$-BN-ELU + MP($2$)		&$2\times$ Conv($3$, pad-1)-$64$-BN-ELU + MP($2$) \\
Conv($1$, pad-0)-$32$-BN-LINEAR	& Conv($1$, pad-0)-$32$-BN-LINEAR \\
GlobalAveragePooling 					& GlobalAveragePooling \\
\hline
\multicolumn{2}{c}{DCCA Optimization Target} \\
\end{tabular}
\end{center}
\caption{\small Architecture of audio-sheet-music retrieval model:
BN: Batch Normalization, ELU: Exponential Linear Unit,
MP: Max Pooling, Conv($3$, pad-1)-$16$: $3 \times 3$ convolution, 16 feature
maps and padding 1}
\label{tab:model_architecture}
\end{table*}

\subsection{Experimental Results}
In the following we provide first quantitative results of our approach.
In terms of evaluation measures we follow the literature \cite{Yan2015DeepCorr}
and report the median rank ($MR$) as well as the $R@k$ rates for both sheet-to-audio
as well as audio-to-sheet retrieval.
The $R@k$ rate (high ist better) is the fraction of queries which have the correct
corresponding snippet in the first $k$ retrieval results.
The $MR$ is the median position (low is better) of the target in a similarity ordered list
of all available snippets in the database.
Table \ref{result-table} provides a summary of our resutls.
For validation and train set we only consider the first 16000 examples
to allow for a direct comparison of the $R@k$ rates with the test set
(to investigate overfitting).
\begin{table}[t]
  \caption{Cross-modality retrieval results on Nottingham dataset}
  \label{result-table}
  \centering
  \begin{tabular}{ccccc|cccc}
    \toprule
    & \multicolumn{4}{c}{Audio-to-Sheet} & \multicolumn{4}{c}{Sheet-to-Audio} \\
    \midrule
    set     & R@1	& R@5	& R@10	& MR		& R@1	& R@5	& R@10	& MR \\
    \midrule
    train (16000)   	& 82.7	& 97.5	& 98.6	& 1		& 81.0	& 97.2	& 98.7	& 1 \\
    valid (16000)  	& 42.0	& 86.6	& 93.1	& 2		& 44.2	& 85.2	& 92.5	& 2 \\
    test (16000)   	& 43.1	& 89.1	& 94.2	& 2		& 45.5	& 87.8	& 93.7	& 2 \\
    \bottomrule
  \end{tabular}
\end{table}
When given an audio snippet from the test set the median rank MR
of the corresponding image snippet is 2 (out of 16000 possible candidates).
The R@10 rate for audio-to-sheet retrieval is 94.2 \%.
This means in particular that for more than 94 \% of the query audio excerpts
the correct sheet image snippet is in the top 10 results of the similarity list
comprising 16000 candidates.
We would like to emphasize that these results are on a completely unseen test set.

\begin{figure}[h!]
 \centerline{\includegraphics[width=1.0\textwidth]{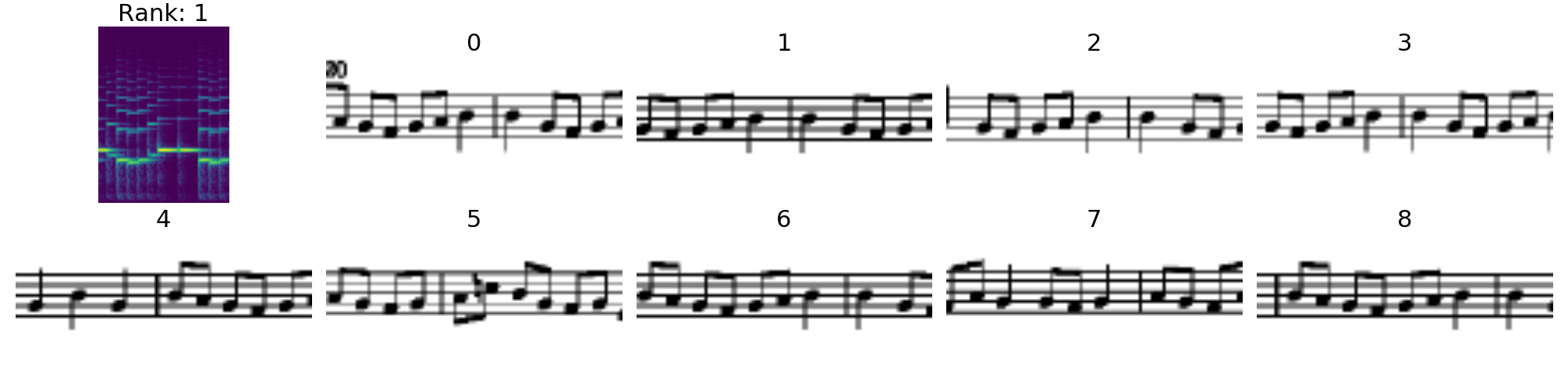}}
 \vspace{-10px}
 \caption{Example of an audio excerpt query and its retrieval results.
 		  The number above the spectrogram is the position of the correct result.
 		  The numbers above the sheet snippets are the ranks of the respective
 		  sheet image snippets in the similarity ordered candidate list.}
 \label{fig:top9}
\end{figure}

%
Figure \ref{fig:top9} shows an example of an audio-to-sheet query along
with its top 9 retrieval results.
The correct sheet snippet is ranked at position 1 for the present case.
However, a closer look at the retrieved images reveals that 4 of the 9 results
(0, 1, 2, 3) are actually only slightly shifted versions
of the ground truth result at position 1 and can be therefore also considered
as correctly retrieved snippets.
This also explains the large gap between the R@1 and R@5 rates
reported above.

%
%

%% file: conclusion.tex
In this work we presented a method for retrieving snippets of sheet music images
when an audio excerpt is given as a search query, and vice versa.
Our solution is based on DCCA that is simultaneously trained on images and audio
in end-to-end neural network fashion.
Once the model is trained it can be used for cross-modality search
when one of the modalities is provided as a retrieval query.
First results suggest that this is a promising research direction
especially in the context of content-based musical retrieval scenarios.